\documentclass[preprint, prb, showpacs, superscriptaddress]{revtex4-1}

\usepackage{graphicx, epstopdf}
\usepackage{subfigure}
\usepackage{float}

\usepackage{amsmath}
\usepackage{amsfonts}
\usepackage{amssymb}
\usepackage{dcolumn}
\usepackage{bm}
\usepackage{subfigure}
\usepackage{caption}

     \newcommand{\beginsupplement}{%
        \setcounter{table}{0}
        \renewcommand{\thetable}{S\arabic{table}}%
        \setcounter{equation}{0}
        \renewcommand{\theequation}{S\arabic{equation}}%
        \setcounter{figure}{0}
        \renewcommand{\thefigure}{S\arabic{figure}}%
     }

\begin{document}

\title{Topological semimetal in honeycomb lattice LnSI}

\author{Simin Nie}
\affiliation{Wuhan National High Magnetic Field Center and School of Physics, Huazhong University of Science and Technology, Wuhan 430074, China}
\affiliation{Department of Materials Science and Engineering, Stanford University, Stanford, California 94305, USA}
\author{Gang Xu}
 \email{gangxu@hust.edu.cn}
\affiliation{Wuhan National High Magnetic Field Center and School of Physics, Huazhong University of Science and Technology, Wuhan 430074, China}
\affiliation{Department of Physics, Mccullough Building, Stanford University, Stanford, California 94305-4045, USA}
\author{Fritz B. Prinz}
\affiliation{Department of Materials Science and Engineering, Stanford University, Stanford, California 94305, USA}
\author{ Shou-cheng Zhang}
\affiliation{Department of Physics, Mccullough Building, Stanford University, Stanford, California 94305-4045, USA}

\date{\today}

\begin{abstract}
Recognized as elementary particles in the standard model, Weyl fermions in condensed matter have received growing attention. However, most of the previously reported Weyl semimetals exhibit rather complicated electronic structures that, in turn, may have raised questions regarding the underlying physics. Here, we report for the first time promising topological phases that can be realized in specific honeycomb lattices, including ideal Weyl semimetal structures, 3D strong topological insulators, and nodal-line semimetal configurations. In particular, we highlight a novel semimetal featuring both Weyl nodes and nodal lines. Guided by this model, we demonstrated that GdSI the long perceived ideal Weyl semimetal has two pairs of Weyl nodes residing at the Fermi level, and that LuSI (YSI) is a 3D strong topological insulator with the right-handed helical surface states. Our work provides a new mechanism to study topological semimetals, and proposes a platform towards exploring the physics of Weyl semimetals as well as related device designs.

\end{abstract}

\maketitle

Weyl fermions (WFs) play a key role in quantum field theory as elementary particles\cite{weyl1929elektron}.
While their existence remains elusive in high energy physics, the realization of WFs in condensed
matter\cite{murakami2007phase,wan2011topological,burkov2011weyl,xu2011chern,soluyanov2015type,
sun2015prediction,liang2016electronic,wang2016observation,deng2016experimental} has attracted considerable interest during
the last few years. In a three dimensional (3D) solid, the low energy excitation of the non-degenerate linearly dispersive band
crossing exactly satisfies the Weyl equation. Such band crossing is named Weyl node (WN), and such a solid is known as the Weyl
semimetal (WSM). According to the Nielsen-Ninomiya theorem\cite{nielsen1983adler}, WNs carrying opposite chiralities must appear
in pairs, between which Fermi arcs can exist at the crystal boundary as a hallmark of the WSMs. Another novel physics property of WSMs is
the chiral anomaly\cite{huang2015observation,zhang2016signatures}, which can result in negative magnetoresistance (NMR)\cite{son2013chiral,arnold2016negative}, nonlocal
electrical transport\cite{parameswaran2014probing} and anomaly phonon-electron coupling\cite{song2016detecting,rinkel2016signatures} \emph{etc.}

Recently, WNs and Fermi arcs were predicted and observed in the TaAs family of compounds\cite{weng2015weyl,huang2015weyl,
lv2015experimental,lv2015observation,xu2015discovery,xu2016observation,yang2015weyl,xu2015discovery2,lv2015observation2}, in which,
up to 24 WNs, as well as many trivial hole and electron Fermi pockets coexist around the Fermi level. Such complicated electronic
structures lead to many debates on the spectroscopic and transport properties, especially the origin of the NMR observed in the TaAs
family. Thus, it is desirable to find the ideal WSMs with less pairs of WNs residing at the Fermi level only.

In this work, we study a special 3D honeycomb lattice model with inversion symmetry broken, and demonstrate that fruitful
topological non-trivial states can be realized in such system, including ideal WSM\cite{ruan2016symmetry,ruan2016ideal}, 3D strong
topological insulator (TI)\cite{hasan2010colloquium,qi2011topological}, nodal-line semimetal\cite{burkov2011topological,weng2015topological,
kim2015dirac,yu2015topological,wang2016body}, and a novel semimetal phase consisting of WNs and nodal lines,
which is discussed for the first time in condensed matters. This model paves a new way to explore the topological materials\cite{tang2016dirac,murakami2016emergence,weng2016topological,xu2016topological,
burkov2016topological,xu2015intrinsic,nie2015quantum,xu2015quantum,ganeshan2015constructing}, especially the ideal WSM and nodal-line semimetal. Furthermore, based on density functional
theory (DFT) calculations, we demonstrate that Rare earth-Sulfide-Iodide LnSI (Ln $=$ Lu, Y and Gd) satisfy this model well, among which LuSI
and YSI are 3D strong TIs with unusual surface states of the right-handed spin texture, and GdSI is the long-pursued ideal WSM with only
2 pairs of WNs crossing the Fermi level. Two very long Fermi arcs exist on the (010) surface of GdSI, which is easily confirmed by the
ARPES experiment. Such ideal WSM phase in GdSI provides great facility for research of the chiral anomaly physics, as well as the
device design based on WSMs.

\section*{Results}

\textbf{Model analysis.} Our tight-binding (TB) model is built on an A-A stacked honeycomb lattice containing two inequivalent sublattices with $|p_z\rangle$
orbitals ($j_z=\pm\frac{1}{2}$) occupied on A-sublattice located at (0, 0, 0), and $|d_{z^2}\rangle$ orbitals ($j_z=\pm\frac{1}{2}$)
occupied on B-sublattice located at ($\frac{1}{3}$, $\frac{2}{3}$, 0), as shown in Figure 1(a), in which only threefold rotation around
$z$-axis ($C_3$), mirror symmetry with respect to $xy$-plane ($M_z$), as well as time reversal symmetry ($\mathcal{T}$) are preserved.
Under the symmetry restrictions, the TB Hamiltonian up to the next-nearest (NN) intralayer and interlayer hoppings takes the form
\begin{eqnarray}
	&&H=H_A+H_B+H_{AB} \nonumber\\
    &&H_{\mu}=\sum_{i}^{\alpha}t_{\mu}^1C^{+}_{\mu\alpha}(i)C_{\mu\alpha}(i)+\sum_{\langle\langle ij\rangle\rangle_{intra}}^{\alpha}t_{\mu}^2C^{+}_{\mu\alpha}(i)C_{\mu\alpha}(j)+\sum_{\langle ij\rangle_{inter}}^{\alpha}t_{\mu}^3C^{+}_{\mu\alpha}(i)C_{\mu\alpha}(j) \\
    &&H_{AB}=\sum_{\langle\langle ij\rangle\rangle_{inter}}^{\alpha}[r_1C^{+}_{A\alpha}(i)C_{B\alpha}(j)+h.c.]+\sum_{\langle ij\rangle_{intra}}^{\alpha\ne\beta}[\lambda_1 C^{+}_{A\alpha}(i)C_{B\beta}(j)+h.c.] \nonumber
\label{HH}
\end{eqnarray}
where $\mu=\text{A}$, B labels the sublattice. $\alpha$, $\beta = \uparrow$, $\downarrow$ label the spin. $C^{+}_{A\alpha}(i)$ ($C^+_{B\alpha}(i) $)
creates a spin $\alpha$ electron in $p_z$ ($d_{z^2}$) orbital of A (B) sublattice at site $i$. The first, second and third terms in $H_\mu$
are on-site energy, NN intralayer hopping and nearest interlayer hopping, respectively. The first term in $H_{AB}$ means NN interlayer hopping,
while the second term is the nearest intralayer hopping induced by the spin-orbit coupling (SOC) interaction.
More detailed definitions of parameters in Equation (1) can be found in Figure 1(a) and Supplementary Section 1.

Compared with the Kane-Mele model\cite{kane2005quantum,kane2005z}, there are three obvious differences in our model. Firstly, our model is
based on a 3D system, which is a necessary condition to realize the WSM. Secondly, the nearest intralayer hopping between the same spin is
forbidden due to the restriction of $M_z$ symmetry. Thus, it is that the nearest intralayer SOC ($\lambda_1$), rather than the NN SOC in the
Kane-Mele model, plays a crucial role for the band gap opening in the $k_z=0$ plane. Finally, inversion symmetry is broken in our model.
Due to the Rashba effect, all bands are split into two branches, which can be distinguished by the eigenvalue of $M_z$, \emph{i.e.}, $m_z=\pm i$
as shown in Figure 1 by dashed ($m_z=i$) and dotted lines ($m_z=-i$). Accordingly, we can define two different splitting configurations:
Configuration I, $p_z$ and $d_{z^2}$ orbitals have the same Rashba splitting as shown in Figure 1(b) and Figure 1(g); Configuration II, $p_z$ and
$d_{z^2}$ orbitals have opposite Rashba splitting as shown in Figure 1(c), 1(e) and 1(h). As we will show below, different Rashba splitting
configurations would lead to different topological states.

For Configuration I, we first study the case that bands only invert with each other around the $\Gamma$ point (named as Case1). For this case,
$p_z$ and $d_{z^2}$ orbitals with the same $m_z$ cross each other at the Fermi level in the $k_z=0$ plane; then re-open a topological
non-trivial insulating gap due to the nearest intralayer SOC ($\lambda_1$) as shown in Figure 1(b); which means a 3D strong TI phase is achieved.
If the band inversion keeps increasing, and all bands are inverted at the $\text{K}$ ($\text{K}'$) point (named as Case2), two pairs of unstable
double-Weyl points ($|C|=2$) should be realized on the H$-$K$-$(-H) and $\text{H}'-\text{K}'-$(-$\text{H}'$) lines as shown in Figure 1(g). The
realization of such double-Weyl points can be understood as following: without loss of generality, we choose A (0,0,0) as the rotation center and define $\hat{R_3^z}=e^{-i\frac{2\pi}{3}\hat{J}_{z}}$ with $\hat{J}_{z}=\hat{L}_{z}+\hat{S}_{z}$, where $\hat{L}_{z}$ and $\hat{S}_{z}$ are the $z$-component of the angular momentum operator and spin operator, respectively. Then we get $\hat{R_3^z}|d^{\{\frac{1}{3}\frac{2}{3}0\}}_{z^2},j_z\rangle_K=e^{-i\frac{2\pi}{3}j_z}|d^{\{\frac{1}{3}-\frac{1}{3}0\}}_{z^2},
j_z\rangle_K=e^{-i\frac{2\pi}{3}j_z}e^{i\frac{2\pi}{3}}|d^{\{\frac{1}{3}\frac{2}{3}0\}}_{z^2},j_z\rangle_K=
e^{-i\frac{2\pi}{3}(j_z-1)}|d^{\{\frac{1}{3}\frac{2}{3}0\}}_{z^2},j_z\rangle_K$, where $K=(-\frac{1}{3},\frac{2}{3},0)$ is defined with respect to the reciprocal lattice vectors.
This means that the effective $j_z$ for the $d_{z^2}$ bands at K point have to decrease by 1, namely become $-\frac{1}{2}$ ($|d_{z^2}\uparrow\rangle$) and $-\frac{3}{2}$($|d_{z^2}\downarrow\rangle$), respectively. Meanwhile, the effective
$j_z^K$ of the $p_z$ bands located at A site do not change at all. As a result, the band crossing between $|j_z=-\frac{3}{2}\rangle$ and
$|j_z=\frac{1}{2}\rangle$ on the H$-$K line should give rise to one double-Weyl point yielding to the requirement that chiral charge $|C|$ equals
to $\Delta j_z=2$. We emphasize that such type of effective $j_{z}$ jumping on the high-symmetry-line,
which is studied for the first time, provides a new mechanism for the exploration of the topological semimetals.

As discussed in Ref. [5], each double-Weyl point has quadratic in-plane (along $k_x$, $k_y$) dispersion and linear out-plane ($k_z$) dispersion.
However, different than HgCr$_2$Se$_4$ with $C_4$ symmetry~\cite{xu2011chern}, the double-Weyl point (\emph{e.g.} $C=2$) in $C_3$ symmetric system is
usually unstable and will split into one negative Weyl point ($C=-1$) and three positive Weyl points ($C=1$)\cite{fang2012multi}
(see details in the Supplementary Section 2 and Figure S1).

For Configuration II, if the bands only invert around the $\Gamma$ point, \emph{i.e.} Case1, it is that the opposite $m_z$ bands cross each other
at the Fermi level in the $k_z=0$ plane as shown in Figure 1(c). No interactions can open band gaps for this case due to the $M_z$ symmetry
protection. Therefore, the system becomes a nodal-line semimetal with two nodal lines circled around the $\Gamma$ point as shown in Figure 1(d).
Given that most proposed nodal-line semimetals exist only by neglecting the effect of SOC\cite{weng2015topological,
kim2015dirac,yu2015topological,wang2016body}, our finding paves a new way for exploration of the
SOC included nodal-line semimetal.

Next, we would like to study the topological states realized for Case2 band inversion with Configuration II
Rashba splitting. For this case, owing to the decrease of $j_z^K(d_{z^2})$ by 1 and the requirement of effective $j_z$ jumping, the system
becomes an ideal WSM phase, in which 4 pairs of linearly dispersive WNs emerge on the H$-$K$-$(-H) and $\text{H}'-\text{K}'-$(-$\text{H}'$) lines
as shown in Figure 1(h), while all the nodal lines are eliminated. More interestingly, a novel semimetal coexisting of both WNs and nodal lines
can be realized in a specific parameter region between Case1 and Case2. In this case, one band inversion crossing occurs on the H$-$K
($\text{H}'-\text{K}'$) line, while the other one is still limited in the $k_z=0$ plane (named as Case3), as shown in Figure 1(e), in which
the left crossing on H$-$K line gives rise a linearly dispersive WN as illustrated for Case2, while the right crossing in the $k_z=0$ plane is
still protected by the $M_z$ symmetry and forms a nodal line around the K point as explained for Case1. As a result, two nodal lines circled
around K and $\text{K}'$ respectively and two pairs of WNs located on the  H$-$K$-$(-H) and $\text{H}'-\text{K}'-$(-$\text{H}'$) lines can be
found in this new topological semimetal as illustrated in Figure 1(f), which is discussed for the first time in condensed matters.

\textbf{Material realization.} Guided by this new model and clear picture, we find a class of topological materials LnSI (Ln $=$ Lu, Y and Gd), among which LuSI and YSI are
3D strong TIs, and GdSI is the long-pursuing ideal WSM with only 2 pairs of WNs crossing the Fermi level. As shown in Figure S2(a, b), LnSI crystallize
in the space group $P\bar{6}$ \cite{GdSIexp,LuSIexp}(same point group as our model), in which Ln atoms (silver-white) and S atoms (yellow) locate in the $z=0$ plane
and form a honeycomb lattice, and I atoms (purple) intercalate between two LnS layers. Our DFT calculations indicate that the low energy bands
near the Fermi level are mainly contributed from the $p_z$ orbitals of S atoms and the $d_{z^2}$ orbitals of the Ln atoms (see the projected
density of states (PDOS) and fatted band analyses shown in Figure S3). In particular, even though there are 4 S atoms and 4 Ln atoms in one unit
cell, only one pair of $p_z$-type molecular orbital $|P_2\rangle$ with $j_z=\pm\frac{1}{2}$ and one pair of $d_{z^2}$-type molecular orbital
$|D_2\rangle$ with $j_z=\pm\frac{1}{2}$ dominate and invert with each other at the Fermi level, owing to the chemical bonding and crystal field
effects. Therefore, our TB model discussed above can be properly applied to LnSI crystal, and capture its essential topological properties
effectively. Detailed evolution from the atomic orbitals to the molecular orbitals is addressed in the Supplementary Section 5.

Since LuSI and YSI have almost the same results, we choose LuSI as an example in the following demonstration. The calculated band structures of
LuSI by the generalized gradient approximation (GGA) and GGA+SOC are shown in Figure 2(a), 2(b) and S3(b), respectively,
which show a very deep band inversion between $p_z$-type $|P_2\rangle$ bands and $d_{z^2}$-type bands $|D_2\rangle$ happens at the $\Gamma$ point.
If we exclude the SOC interaction, this band inversion will result in a nodal line centered around $\Gamma$ point in the $k_z$ = 0 plane, as shown
in the inset of Figure 2(a) by the GGA calculations. When the SOC is considered, we have calculated the eigenvalues of the mirror symmetry $M_z$ for
the $|P_2\rangle$ and $|D_2\rangle$ bands. The calculated results show that  $|P_2\rangle$ and $|D_2\rangle$ bands have the same Rashba splitting
in LuSI, \emph{i.e.}, LuSI conforms to Case1 band inversion of Configuration I splitting. So that GGA+SOC calculations for LuSI show a 32 meV
topological non-trivial band gap as shown in the Figure 2(b). In order to check its topological properties, we have carried out the calculations of
surface states for LuSI by constructing the Green's functions\cite{sancho1984quick,sancho1985highly} based on the maximally localized Wannier function (MLWF) method\cite{marzari2012maximally}. The calculated
results in Figure 3(a) indicate that there is a surface Dirac cone in the band gap connecting the occupied and unoccupied bulk states at the $\bar{\Gamma}$
point on the (001) face of LuSI, which confirms LuSI is a 3D strong TI clearly. It is worth noting that, different from most 3D TIs with the
left-handed helical Dirac cones, the surface states of LuSI exhibits a right-handed helicity of the spin-momentum locking, as shown in Figure 3(b),
which indicates a negative SOC in LuSI\cite{sheng2014topological}.

In the next step, we study the topological properties of GdSI. Considering that the $f$ orbitals of Gd are partially occupied, GdSI is very likely to
stabilize in a magnetic phase. We have calculated five different magnetic configurations for GdSI by the GGA+SOC, including the ferromagnetic (FM),
three collinear antiferromagnetic configurations (AFM1-AFM3), and one non-collinear collinear antiferromagnetic configuration (AFM4) as shown in
Figure S5. The calculated total energies and moments are summarized in Table S1, which demonstrates that all magnetic states are lower than the
non-magnetic (NM) state about 24 eV/ u.c., and the AFM4 configuration is the most stable one, further lowering the total energy about 10-20 meV than
the other collinear magnetic states. This is because that AFM4 configuration has eliminated the frustrations as much as possible, and it agrees
with the $2\times2$ reconstruction of the crystal mostly \cite{GdSIexp,LuSIexp}.

In order to deal with the correlation effect of the $f$ electrons, we have performed the GGA+Hubbard U (GGA+U) calculations on GdSI. The GGA+U and
GGA+U+SOC band structures of AFM4 are plotted in Figure 2(c) and 2(d), respectively, which show a similar dispersion to LuSI at a quick glance.
However, after a meticulous analysis, we find three substantial differences from LuSI. Firstly, our calculations indicate that $|P_2\rangle$ and
$|D_2\rangle$ bands in GdSI take the opposite Rashba splitting Configuration II. Secondly, band inversion in GdSI not only exists at the $\Gamma$
point, but also happens at the K (K$'$) point, \emph{i.e.}, GdSI belongs to band inversion Case3. Finally and most importantly, both time reversal
symmetry $\mathcal{T}$ and mirror symmetry $M_z$ are broken in the ground state AFM4 of GdSI as a result of the non-collinear magnetic configuration.
So that the band crossing in the $k_z=0$ plane has lost the $M_z$ protection, and opens a gap because $m_z$ is not a good quantum number again. Based
on this symmetry analysis and as will be shown below, GdSI becomes an ideal WSM with two pairs of WNs originating from the band crossing occurring on the H$-$K
($\text{H}'-\text{K}'$) line, though GdSI is categorized to Case3 of Configuration II

For describing GdSI's band structures and topological properties accurately, a Zeeman splitting term $H_z=\left( \begin{smallmatrix}
              t^4_A & 0 \\ 0 & -t^4_B
           \end{smallmatrix}  \right)\bigotimes
\left( \begin{smallmatrix}
              1 & 0 \\ 0 & -1
           \end{smallmatrix}  \right)$  that breaks the time reversal symmetry $\mathcal{T}$, and a nearest intralayer hopping $r_2=\langle
 p^{\{000\}}_z\uparrow|H|d^{\{\frac{1}{3}\frac{2}{3}0\}}_{z^2}\uparrow\rangle$ that breaks the $M_z$ symmetry are added to the TB model Eq. (1).
The explicit form of this new Hamiltonian and the fitted parameters for GdSI are described in the Supplementary Section 7. The fitted band
structures (red dots) are plotted together with the GGA+U+SOC bands (blue lines) in Figure 2(d), which demonstrates that the effective model
reproduces the DFT calculations quantitatively well. Based on this effective TB model and the fitted parameters, we have calculated the chiral
charges for the WNs located above the $k_z=0$ plane, respectively, and plotted their evolution\cite{yu2011equivalent} in Fig 2(e), which manifests that the charge
center for the WN located on the top of K point shifts downward (red dots), indicating the Chern number $C=-1$, while the charge center for
the WN located located on the top of K$'$ point shifts upward (blue dots), corresponding to $C=1$. The WNs distribution in the $k_z > 0$
Brillouin zone (BZ) is summarized in Figure 2(f), and we find their counterparts at the same $k_x$, $k_y$ but opposite $k_z$, because the
inverted bands are approximately symmetrical around K (K$'$) point as shown in Figure S6, in spite of the $M_z$ symmetry breaking in GdSI. Such
conclusion is completely consistent with our DFT calculations, which indicate that GdSI holds only two pairs of WNs located at
($-\frac{1}{3}$, $\frac{2}{3}$, $\pm 0.023$) and ($\frac{1}{3}$, $-\frac{2}{3}$, $\pm 0.021$) crossing the Fermi level. Note that the small
difference between the dispersions around K point and K$'$ point is induced by the time reversal symmetry breaking.

Based on the effective TB model, we have calculated the (001) surface states and Fermi arcs on the (010) surface of GdSI, and plotted them in
Figure 3(c) and 3(d), respectively. The (001) surface state calculation exhibits a clear band touching at the $\bar{\text{K}}$ point and
Fermi level, indicating that GdSI is an ideal WSM. However, because two bulk WNs carrying opposite chiralities and same in-plane coordinates are
projected to the same point, no Fermi arc can be found on the (001) face, as shown in Figure S7(a). In contrast, as shown in Figure 3(d), two long Fermi arcs connecting
the opposite WNs exist on the (010) face unambiguously, which provides great facility for the ARPES experiment to confirm its topological
properties.

\section*{Conclusion}
In summary, we have studied a specific 3D honeycomb model, in which fruitful topological phases can be realized, including ideal WSM, 3D strong TI,
nodal-line semimetal, and the novel semimetal consisting of both WNs and nodal lines, suggesting a new mechanism for exploring the topological semimetals.
Guided by this model, our DFT calculations predict that LuSI and YSI are 3D strong TIs with unusual right-handed helical Dirac cones, and GdSI, which
stabilized in a non-collinear AFM states, is the long-pursuing ideal WSM with two pairs of WNs residing at the Fermi level. Furthermore, there are two very
long Fermi arcs on the (010) surface of GdSI, which are well-suited for the ARPES measurement. Such ideal WSM phase in GdSI provides a good platform to
study the physics of the chiral anomaly, and great facility for the applications of the WSMs

{\fontsize{15}{15}\selectfont\textbf{Methods}}

The DFT calculations are performed by the projector augmented wave method implemented in
Vienna $ab~ initio$ simulation package (VASP)\cite{kresse1996_1,kresse1996_2}. The cut-off energy is 500 eV. GGA of
Perdew-Burke-Ernzerhof type\cite{Perdew1996} is used to treat with the exchange and correlation potential. SOC is taken
into account self-consistently. The $k$-points sampling grid of the BZ is 5 $\times$ 5 $\times$ 11. The GGA+U
scheme\cite{liechtenstein1995density} is use to induce an effective on-site Coulomb potential of 6.0 eV for the $f$
orbitals of $\text{Gd}$. MLWFs have been generated to construct the TB Hamiltonians of semi-infinite
sample\cite{marzari2012maximally}. The projected surface states are obtained from the TB Hamiltonians by using an iterative
method\cite{sancho1984quick,sancho1985highly}.

{\fontsize{15}{15}\selectfont\textbf{Acknowledgements}}

We thank Biao Lian and Zhida Song for useful discussions. G. X. is supported by the National Thousand-Young-Talents Program and the NSFC.
F. B. P. and S. N. are supported by Stanford Energy 3.0. S.-C. Z. is supported by the U.S. Department of Energy, Office of Basic Energy Sciences, Division of
Materials Sciences and Engineering under Contract No.~DE-AC02-76SF00515, by FAME, one of six centers of STARnet, a Semiconductor Research
Corporation program sponsored by MARCO and DARPA.

{\fontsize{15}{15}\selectfont\textbf{Author contributions}}

S. N., G. X. and S.-C. Z conceived and designed the project. S. N. and G. X performed all the DFT calculations and
theoretical analysis. All authors contributed to the manuscript writing.

{\fontsize{15}{15}\selectfont\textbf{Competing financial interests}}

The authors declare no competing financial interests.

{\fontsize{15}{15}\selectfont\textbf{Correspondence}}
Correspondence and requests for materials should be addressed to G. Xu~(email: gangxu@hust.edu.cn).

\newpage
\renewcommand\figurename{\textbf{Figure}}
\begin{figure}[t]
\includegraphics[clip,scale=0.4, angle=0]{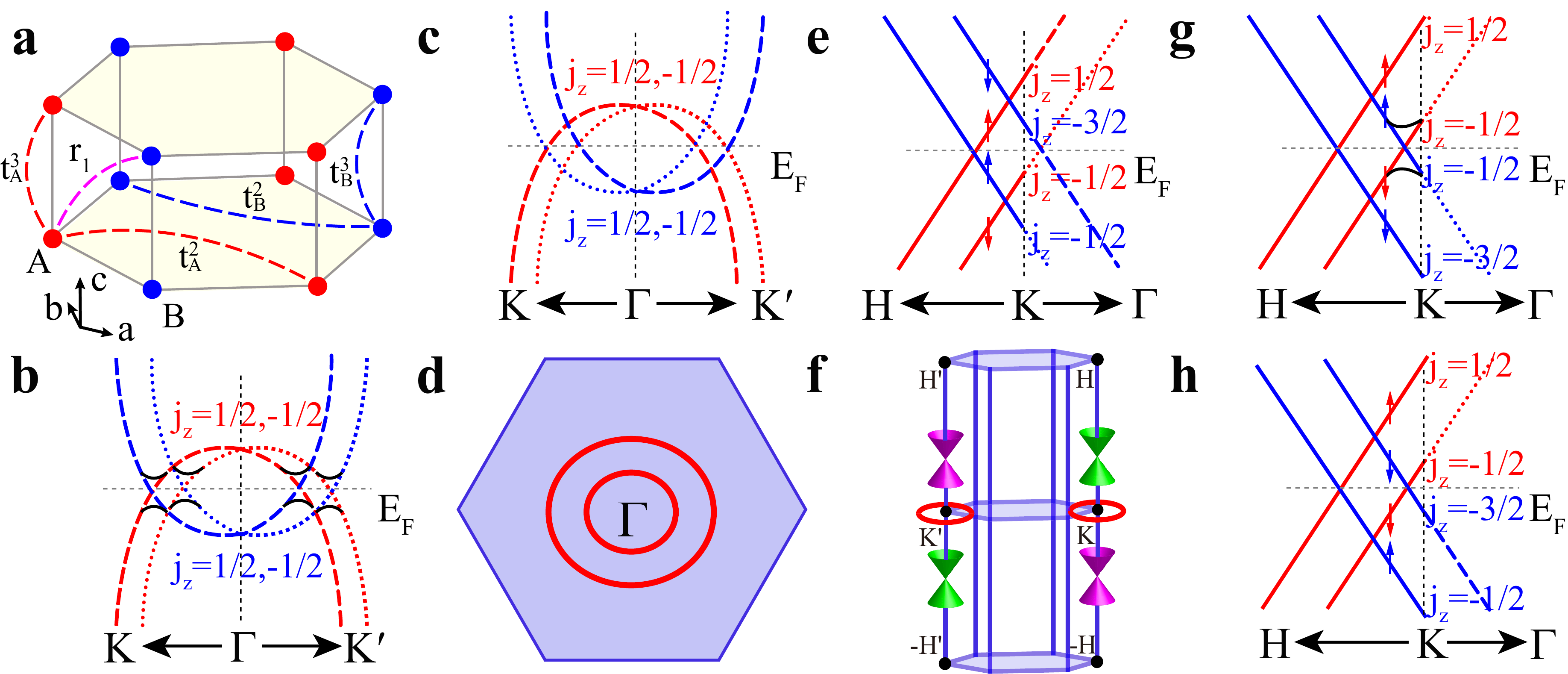}
\caption{~$\boldsymbol{|}$ \textbf{Schematic of topological phases.} \textbf{(a)} The A-A stacked honeycomb lattice and hopping parameters, where A-sublattice (red dots) is occupied by
$p_z$ orbital, while B-sublattice (blue dots) is occupied by $d_{z^2}$ orbital. \textbf{(b)} 3D strong TI. \textbf{(c)} Nodal-line semimetal with two nodal lines circled around the
$\Gamma$ point as shown in \textbf{d}. \textbf{(e)} Novel topological semimetal coexisting of both WNs and nodal lines as shown in \textbf{f}. \textbf{(g)} Double-Weyl semimetal with two
 pairs of double-Weyl points. \textbf{(h)} Ideal WSM holding 4 pairs of WNs. The red (blue) lines in \textbf{b}, \textbf{c}, \textbf{e}, \textbf{g} and \textbf{h} represent the
 $p_z$-type ($d_{z^2}$-type) bands, and different eigenstates of $M_z$ are distinguished by the dashed ($m_z = i$) and dotted lines ($m_z = -i$), respectively. The effective $j_z$ at
 the $\Gamma$ or K point, as well as the spin direction for each band are also labeled.}
\end{figure}

\newpage
\begin{figure*}[htp]
\center
\includegraphics[clip,scale=0.28, angle=0]{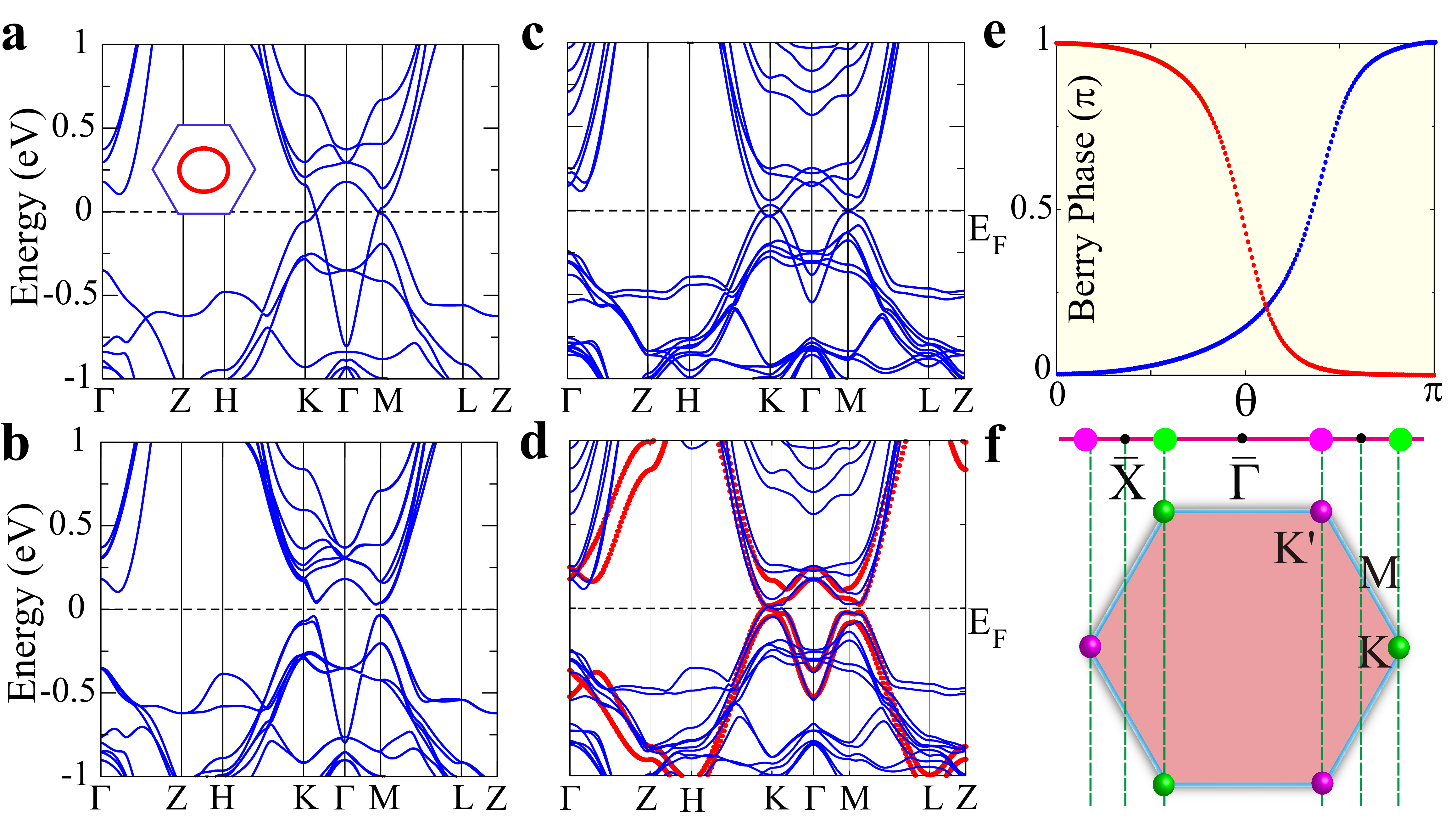}
      \caption{~$\boldsymbol{|}$ \textbf{Band structures and WNs.} \textbf{(a, b)} The band structures of LuSI calculated by GGA and GGA+SOC, respectively. The inset in \textbf{a}
      shows the schematic of the nodal line when SOC is neglected. \textbf{(c, d)} Band structures of the ground state in GdSI calculated by GGA+U and GGA+U+SOC, respectively. The
      red dots in \textbf{d} are the fitted TB results. \textbf{(e)} Chiral charge evolution of the WNs located on K$-$H (red dots) and K$'-$H$'$ (blue dots) of GdSI. (f) Summary of
      the WNs distribution for the $k_z > 0$ BZ of GdSI, where green and magenta balls mean the negative ($C=-1$) and positive ($C=+1$) WNs, respectively. The projections of the WNs on the
      (010) face are shown too.
      }
\label{structure}
\end{figure*}

\newpage\newpage
\newpage\clearpage
\begin{figure}[tp]
\includegraphics[clip,scale=0.17, angle=0]{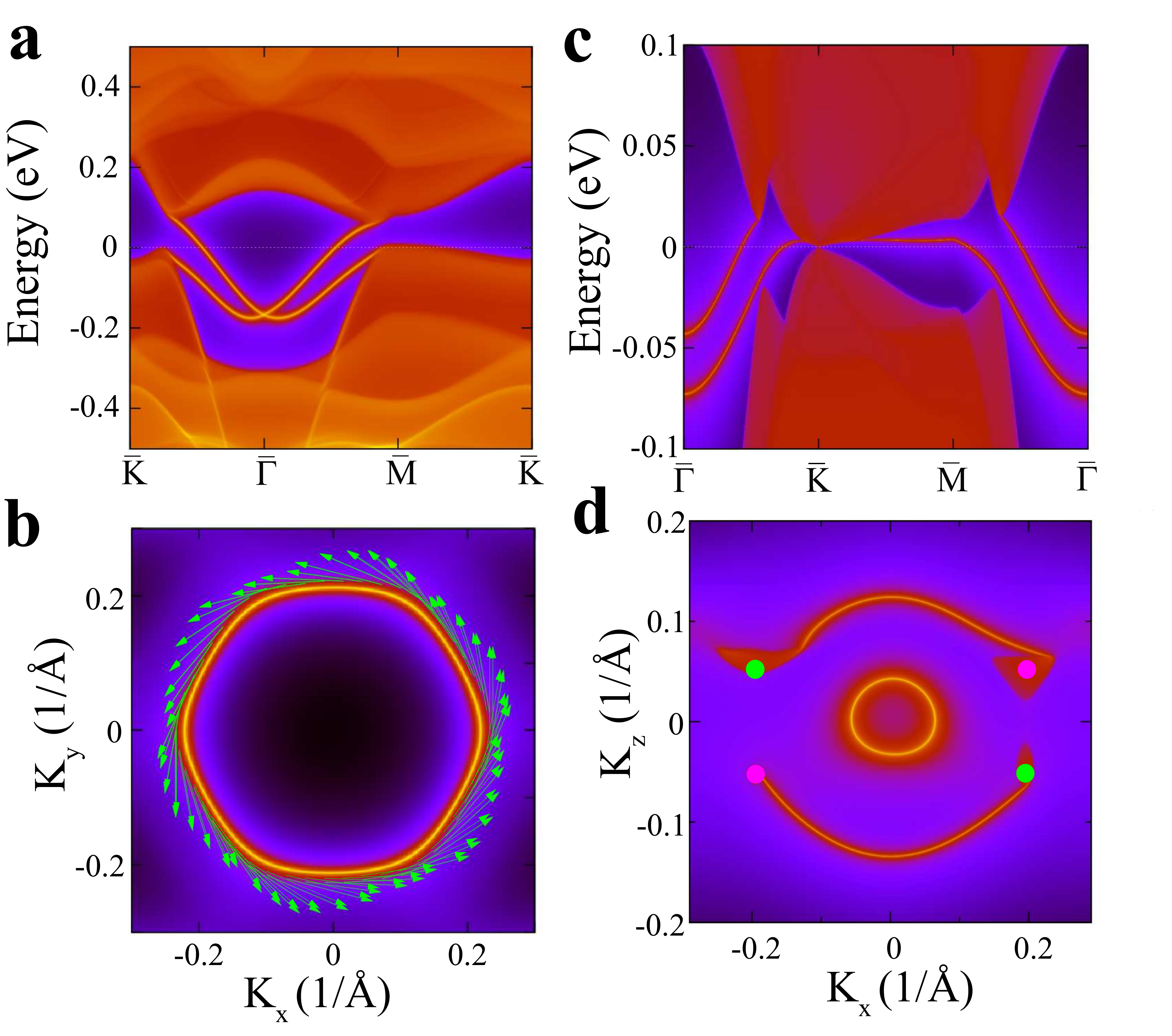}
      \caption{~$\boldsymbol{|}$ \textbf{Surface states and Fermi arcs.} \textbf{(a, c)} Band structures of LuSI and GdSI projected onto the (001) face, respectively.
      \textbf{(b)} Topological surface states and corresponding spin texture on the (001) face of LuSI.
      \textbf{(d)} Fermi arcs in the (010) surface BZ of GdSI.
      }
\label{structure}
\end{figure}

\pagebreak

\newpage
\newpage
\newpage
\clearpage

\beginsupplement
{\fontsize{15}{15}\selectfont\textbf{Supplementary Information}}

\captionsetup[figure]{labelformat={default},labelsep=period,name={Fig.}}

Contents:

\begin{itemize}
  \item { S1. TIGHT-BINDING MODEL IN THE MOMENTUM SPACE}
  \item { S2. DOUBLE-WEYL POINT SPLITTING IN $C_3$ SYMMETRIC SYSTEM}
  \item { S3. CRYSTAL STRUCTURE AND BRILLOUIN ZONE}
  \item { S4. PROJECTED DENSITY OF STATES AND FATTED BANDS}
  \item { S5. MOLECULAR ORBITALS AND BAND EVOLUTION AT THE $\Gamma$ POINT}
  \item { S6. MAGNETIC CONFIGURATIONS AND TOTAL ENERGY CALCULATIONS IN GdSI}
  \item { S7. MIRROR SYMMETRY $M_z$ BREAKING AND FITTED PARAMETERS IN \text{GdSI}}
  \item { S8. SURFACES STATES OF GdSI}
\end{itemize}

\newpage
\section*{S1. TIGHT-BINDING MODEL IN THE MOMENTUM SPACE}

\begin{figure}[ht]
\centering
\includegraphics[width=6.2in]{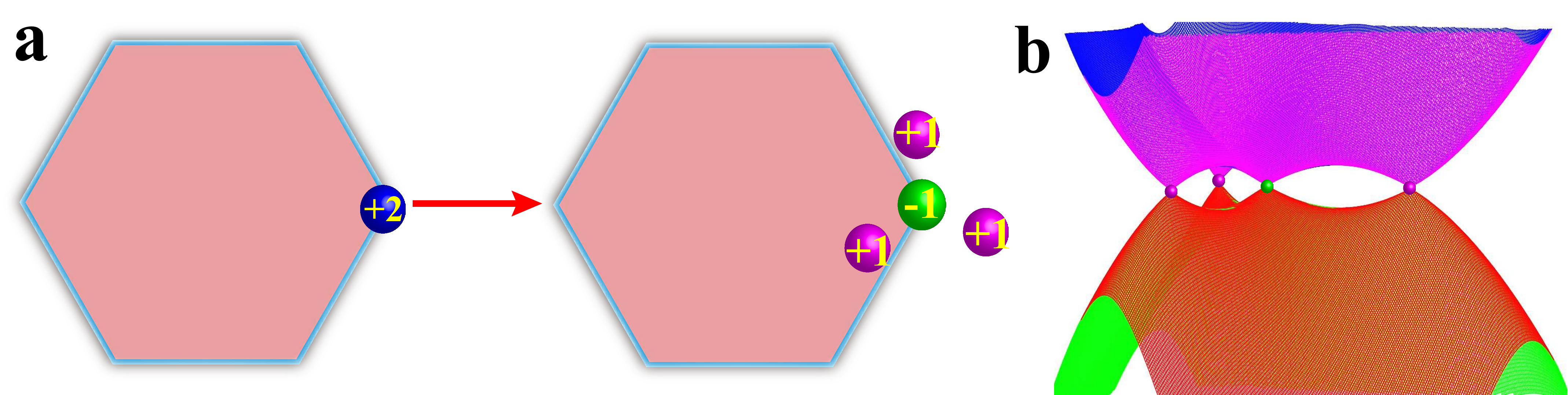}
      \caption{(Color online). (a) Schematic splitting of the double-Weyl point in the  $C_3$ symmetric system.
      Blue, magenta and green balls represent the Weyl points with $C= +2$, +1 and -1, respectively. (b) Energy
      spectrum showing the in-plane ($k_xk_y$-plane) linear dispersion around the negative and three positive
      Weyl points at fixed $k_z$.}
\label{splitting}
\end{figure}

In the momentum space, the Bloch bases for our tight-binding (TB) model can be constructed as
\begin{align}
|\mu\alpha,\mathbf{k}\rangle=\frac{1}{\sqrt{N}}\sum_{\mathbf{R}_i} e^{i\mathbf{k}\cdot(\mathbf{R}_i+\mathbf{r}_{\mu})}\phi_{\alpha}(\mathbf{r}-\mathbf{R}_i-\mathbf{r}_{\mu})
\end{align}
in which the definitions of $\mu$ and $\alpha$ are given in the main text, $\mathbf{k}$ is the crystal
momentum, N is the number of unit cells, $\mathbf{R}_i$ is the lattice vector, $\mathbf{r}_{\mu}$ is
the position of sublattice $\mu$, and $\phi_{\alpha}(\mathbf{r}-\mathbf{R}_i-\mathbf{r}_{\mu})$ is the
atomic orbital wave function. Because only one real spherical harmonic wave function is used for each
sublattice, here we use the sublattice to label the spatial orbital too, which means that
$\phi_{\alpha}(\mathbf{r}-\mathbf{R}_i-\mathbf{r}_{A})$ ($\phi_{\alpha}(\mathbf{r}-\mathbf{R}_i-\mathbf{r}_{B})$)
is the $p_z$ ($d_{z^2}$) orbital with spin $\alpha$ on A(B)-sublattice located at
$\mathbf{R}_i+\mathbf{r}_{A}$ ($\mathbf{R}_i+\mathbf{r}_{B}$).

~\\

By means of the following transformations:
\begin{align}
C^{+}_{\mu\alpha}(i)=\frac{1}{\sqrt{N}}\sum_{\mathbf{k}} e^{i\mathbf{k}\cdot(\mathbf{R}_i+\mathbf{r}_{\mu})}C^{+}_{\mu\alpha}(k) \\
C_{\mu\alpha}(i)=\frac{1}{\sqrt{N}}\sum_{\mathbf{k}} e^{-i\mathbf{k}\cdot(\mathbf{R}_i+\mathbf{r}_{\mu})}C_{\mu\alpha}(k)
\end{align}
the TB Hamiltonian Eq. (1) in the main text can be transferred into the momentum space and written as
the matrix form with basis order $|A \uparrow,\mathbf{k}\rangle$, $|A \downarrow,\mathbf{k}\rangle$,
$|B \uparrow,\mathbf{k}\rangle$ and $|B \downarrow,\mathbf{k}\rangle$.

{\fontsize{10.3}{10}\selectfont
     \begin{align}
        H(\boldsymbol{k}) =
        &\left( \begin{array}{cccc}
            M_A                      &                   0         &  2ir_1Gsin(k_z)          & \lambda_1 D   \\
                                   0    & M_A                     & \lambda_1 F         & 2ir_1Gsin(k_z) \\
                                   *    &                           * &  M_B  &       0            \\
                                   *    &                           * &                0             &   M_B
        \end{array}\right)  \label{H0k} \\
        \nonumber
    \end{align}
}where $M_A=t_A^{1}+2t_A^{2}(cos(k_x)+cos(k_y)+cos(k_{x}+k_{y}))+2t_A^{3}cos(k_{z})$,
$M_B=t_B^{1}+2t_B^{2}(cos(k_x)+cos(k_y)+cos(k_{x}+k_{y}))+2t_B^{3}cos(k_{z})$, $G=e^{i\beta}+e^{i\gamma}+e^{i\eta}$,
$D=e^{i\beta}+e^{-i\frac{2\pi}{3}}e^{i\gamma}+e^{-i\frac{4\pi}{3}}e^{i\eta}$ and
$F=e^{i\beta}+e^{i\frac{2\pi}{3}}e^{i\gamma}+e^{i\frac{4\pi}{3}}e^{i\eta}$
with $\beta=\frac{1}{3}k_x+\frac{2}{3}k_y$, $\gamma=-\frac{2}{3}k_x-\frac{1}{3}k_y$ and $\eta=\frac{1}{3}k_x-\frac{1}{3}k_y$.
As shown in Fig. 1(a) of the main text, we define $t_A^1$ and $t_B^1$ as the on-site energy of A-sublattice and B-sublattice,
respectively; $t_A^2=\langle p^{\{000\}}_z\uparrow|H|p^{\{100\}}_z\uparrow\rangle$ ($t_B^2=\langle d^{\{\frac{1}{3}\frac{2}{3}0\}}_{z^2}\uparrow|H|d^{\{\frac{4}{3}\frac{2}{3}0\}}_{z^2}\uparrow\rangle$) means the next
nearest (NN) intralayer hopping between A-sublattices (B-sublattices) with the same spin;
$t_A^3=\langle p^{\{000\}}_z\uparrow|H|p^{\{001\}}_z\uparrow\rangle$ ($t_B^3=\langle d^{\{\frac{1}{3}\frac{2}{3}0\}}_{z^2}\uparrow|H|d^{\{\frac{1}{3}\frac{2}{3}1\}}_{z^2}\uparrow\rangle$) means the nearest
interlayer hopping between A-sublattices (B-sublattices) with the same spin;
$r_1=\langle p^{\{000\}}_z \uparrow|H|d^{\{\frac{1}{3}\frac{2}{3}1\}}_{z^2}\uparrow\rangle$ means the NN interlayer hopping
of the same spin; and $\lambda_1=\langle p^{\{000\}}_z\uparrow|H|d^{\{\frac{1}{3}\frac{2}{3}0\}}_{z^2}\downarrow\rangle$ means
the nearest intralayer hopping between the opposite spin, which is originated from the spin-orbit coupling (SOC) interaction.
We note that the nearest intralayer hopping between $|p_z\rangle$ and $|d_{z^2}\rangle$ orbitals with the same spin is forbidden
in the honeycomb lattice, when $M_z$ symmetry is preserved.

 \section*{S2. DOUBLE-WEYL POINT SPLITTING IN $C_3$ SYMMETRIC SYSTEM}

As discussed in the main text, two pairs of double-Weyl points ($|C| = 2$) should be realized in the Case2 band inversion of
Configuration I Rashba splitting, due to the $\Delta j_z$ jumping 2. However, such double-Weyl point in the $C_3$ symmetric
system is unstable according to previous analysis\cite{fang2012multi}. As shown in Fig. \ref{splitting}(a), one double-Weyl
point ($C=2$) located on the $\text{H}-\text{K}$ line will split into one negative Weyl point ($C=-1$) located on the $\text{H}-\text{K}$
line and three positive Weyl points ($C=1$) related by $C_3$ symmetry. In Fig. \ref{splitting}(b), we also plot the energy
spectrum around the negative and three positive Weyl points, which shows a linear in-plane ($k_xk_y$-plane) dispersion at
each node point, rather than the quadratic in-plane dispersion for the double-Weyl point as shown in HgCr$_2$Se$_4$ \cite{xu2011chern}.

\section*{S3. CRYSTAL STRUCTURE AND BRILLOUIN ZONE}

\begin{figure}[ht]
\centering
\includegraphics[width=3.9 in]{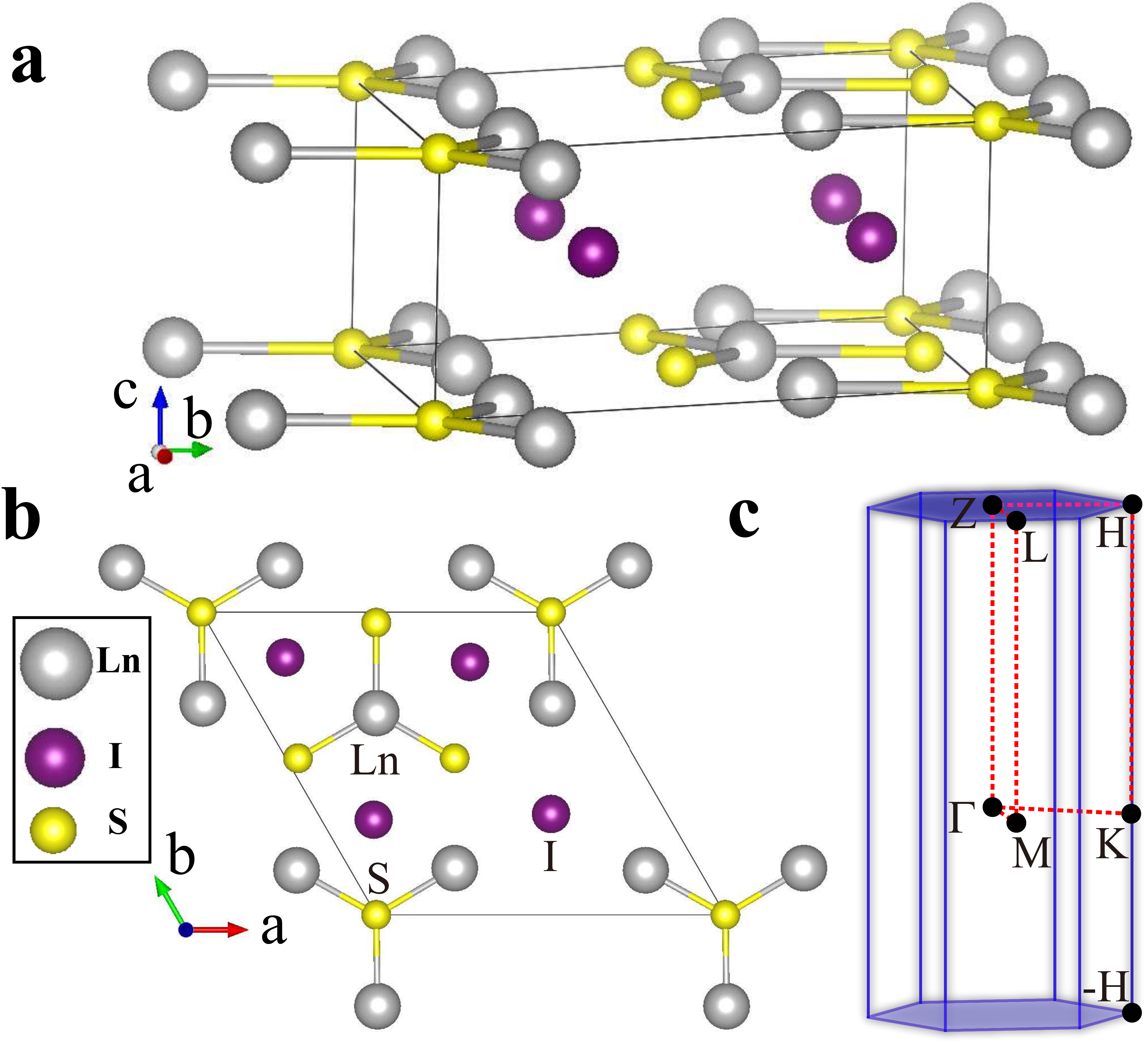}
      \caption{(Color online). Side view (a) and top view (b) of LnSI.
       Silver-white, yellow and purple balls are Ln, S and I atoms, respectively.
       (c) Bulk BZ with high symmetry points (black dots).}
\label{structure}
\end{figure}

The crystal structure and Brillouin zone (BZ) of LnSI (Ln = Lu, Y and Gd) are shown in Fig. \ref{structure}.
Experimentally, LnSI crystallize in the 174 space group with 4 formulas per unit cell \cite{GdSIexp,LuSIexp},
in which each type of atoms can be classified into two different positions. Taking S atom as an example,
one S atom is located at the \emph{1a} site, which is invariant under $C_3$ symmetry, while the other
three S atoms are located at the \emph{3j} sites which are related by $C_3$. As shown in Fig. \ref{structure}(a,b),
even though LnSI suffers a 2 $\times$ 2 reconstruction, Ln and S atoms are still located in the same plane,
and can be taken as a honeycomb lattice roughly. More importantly, our TB model built on the perfect
honeycomb lattice can capture the low energy physics in LnSI very well as we'll analyse later.

\begin{figure*}[ht]
\centering
\includegraphics[width=5.5 in]{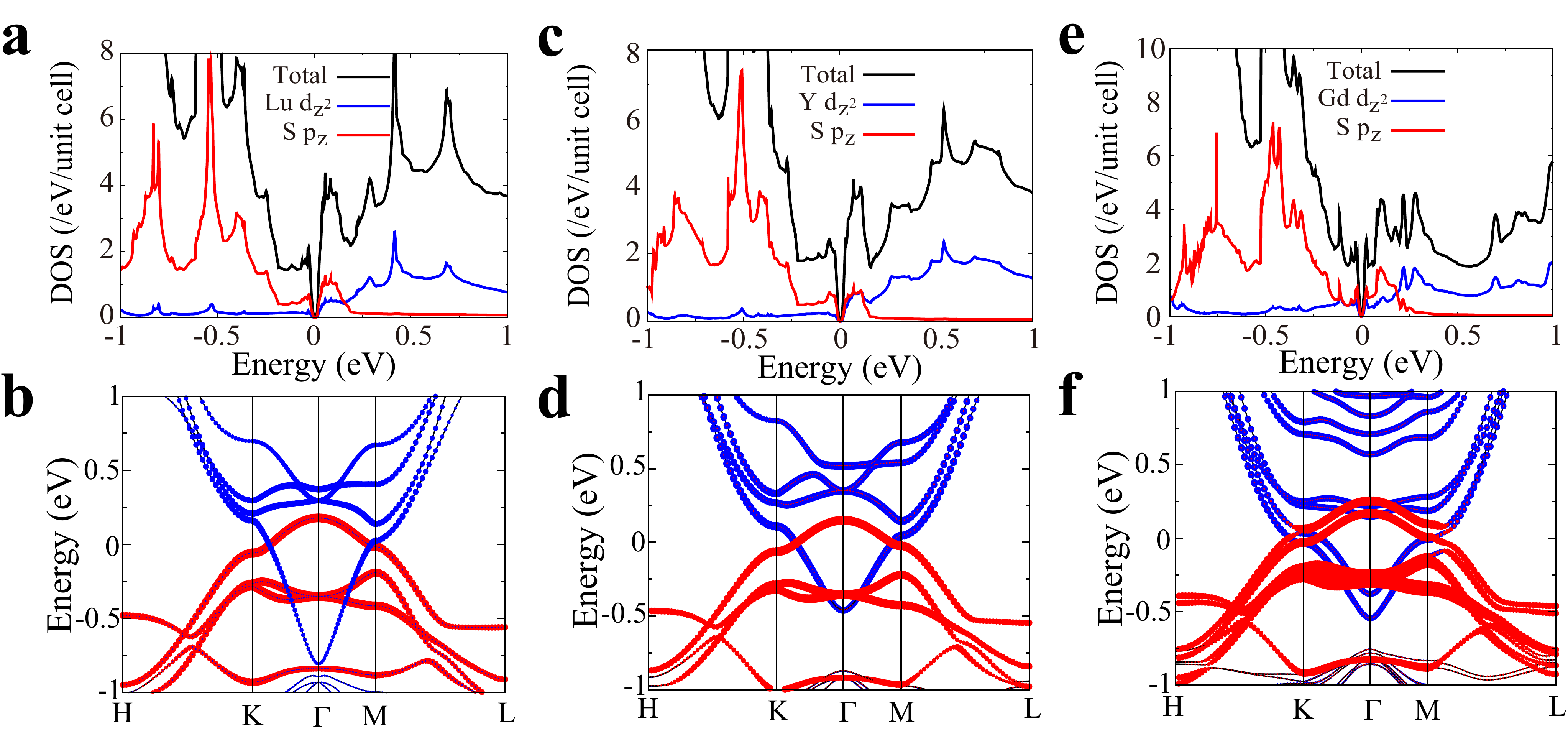}
      \caption{(Color online) (a), (c) and (e) are the GGA (GGA+U for GdSI) calculated TDOS and PDOS of LuSI (NM), YSI (NM) and GdSI (AFM4), respectively.
      (b), (d) and (f) are the corresponding fatted bands of LuSI, YSI and GdSI, respectively. The size of red and blue solid dots in the fatted bands
      represents the weight of S $p_z$ and Ln $d_{z^2}$, respectively. The effect of SOC is excluded in all these calculations}
\label{pdos}
\end{figure*}

\section*{S4. PROJECTED DENSITY OF STATES AND FATTED BANDS}

The total density of states (TDOS) and projected density of states (PDOS) of LnSI calculated by GGA (GGA+U for GdSI) are plotted in the up panel of Fig. \ref{pdos}, which clearly show
that the valence and conduction bands around the Fermi level (0 eV) are dominated by the S $p_z$ and Ln $d_{z^2}$ states. To be specific, TDOS is mainly contributed by S $p_z$ states
from $-1$ to 0 eV, and by Ln $d_{z^2}$ states from 0 to 1 eV. In particular, there exists an obvious weight exchange near the Fermi level in LnSI, indicating a band inversion between S
$p_z$ and Ln $d_{z^2}$ states, which can be presented more clearly by corresponding fatted bands plotted in the down panel of Fig. \ref{pdos}. The size of the red solid dots in Fig.
\ref{pdos} represents the projection of S $p_z$ states, while the size of blue solid dots represents the projection of Ln $d_{z^2}$ states. Consistent with the PDOS results, the fatted
bands in Fig. \ref{pdos} intuitively show that one S $p_z$ band and one Ln $d_{z^2}$ band (ignoring the spin degree) invert with each other around the $\Gamma$ point, which will lead
to some topological non-trivial properties in these materials.

\section*{S5. MOLECULAR ORBITALS AND BAND EVOLUTION AT THE $\Gamma$ POINT}

\begin{figure}[tp]
\centering
\includegraphics[width=3.9 in]{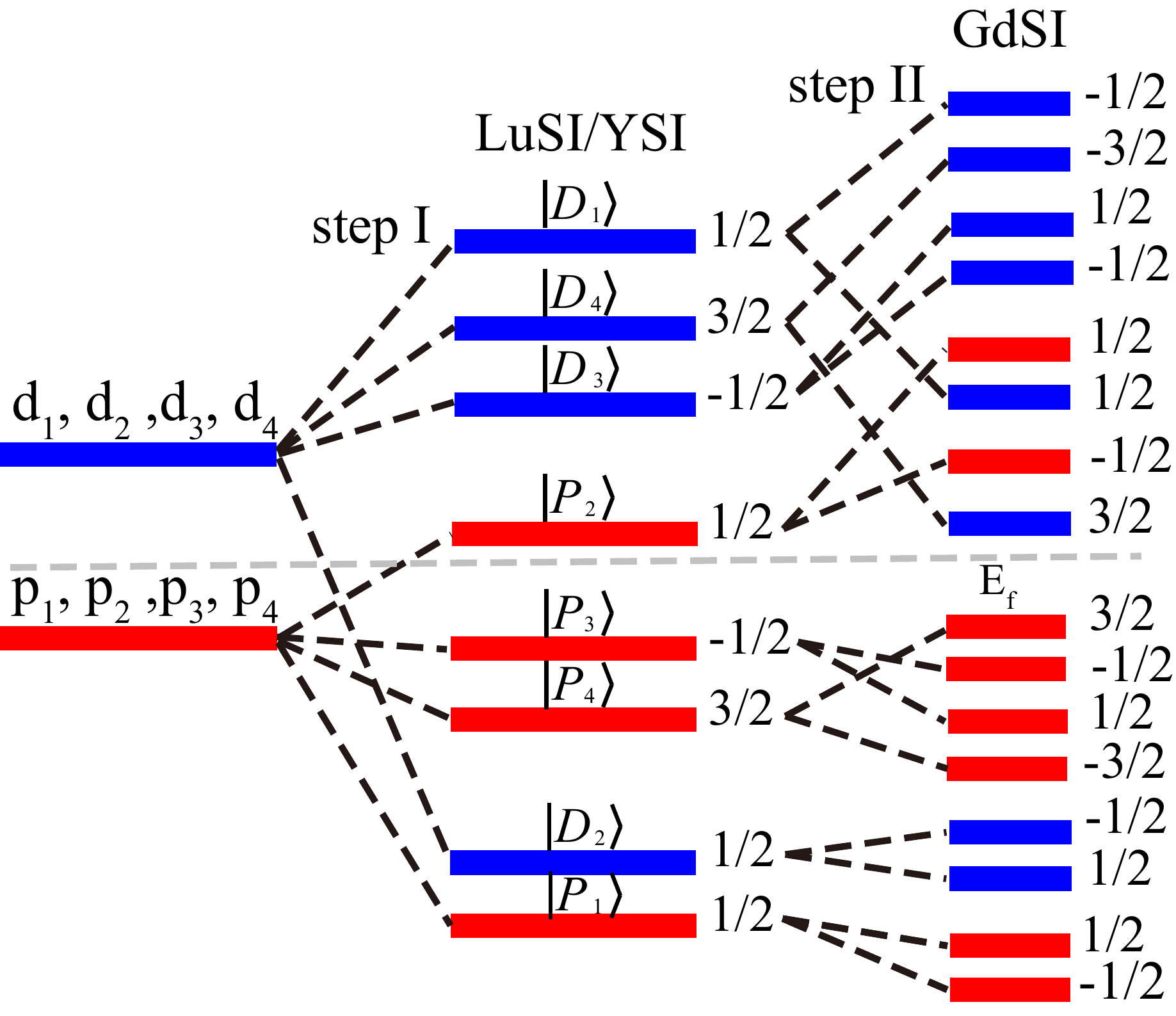}
      \caption{(Color online). Schematic diagram of the band evolution at the $\Gamma$ point in $\text{LnSI}$, starting from the atomic orbitals, $d_{z^2}$ of $\text{Ln}$ and $p_z$ of
      $\text{S}$. The following two steps are required to understand the band orders at the $\Gamma$ point: (I) the chemical bonding and crystal field effect, (II) the exchange coupling
      of the non-collinear magnetic configuration. $E_f$ is the Fermi level (gray dashed line).}
\label{evolution}
\end{figure}

In this section, we would like to demonstrate the formation of the molecular orbitals and the band evolution at $\Gamma$ point in LnSI.
Considering that SOC only plays a role in the band gap opening, we ignore its effect in the following discussion. The schematic diagram
of the band evolution at the $\Gamma$ point in $\text{LnSI}$ is plotted in Fig. \ref{evolution}, where $d_i$ and $p_i$ represent the $d_{z^2}$
orbital of $i$-th $\text{Ln}$ and $p_z$ orbital of $i$-th $\text{S}$, respectively. At step I, we consider the chemical bonding and the
crystal field effects, where $S_z$ is a good quantum number and $M_z$ symmetry is preserved. Therefore, we can focus on the spin up
channel only, and the spin down channel can be obtained easily with the help of time reversal symmetry $\mathcal{T}$. Due to the chemical
bonding and the crystal field effects, eight molecular orbitals can be constructed from the atomic orbitals ($d_i$ and $p_i$) by using the
$C_3$ symmetry. The explicit formulas of the eight molecular orbitals of up spin are listed as following: $|P_1\rangle=|p_1\uparrow\rangle$,
$|P_2\rangle=\frac{1}{\sqrt{3}}(|p_2\uparrow\rangle+|p_3\uparrow\rangle+|p_4\uparrow\rangle)$,
$|P_3\rangle=\frac{1}{\sqrt{3}}(\alpha|p_2\uparrow\rangle+\alpha^2|p_3\uparrow\rangle+|p_4\uparrow\rangle)$,
$|P_4\rangle=\frac{1}{\sqrt{3}}(\alpha^2|p_2\uparrow\rangle+\alpha|p_3\uparrow\rangle+|p_4\uparrow\rangle)$,
$|D_1\rangle=|d_1\uparrow\rangle$,
$|D_2\rangle=\frac{1}{\sqrt{3}}(|d_2\uparrow\rangle+|d_3\uparrow\rangle+|d_4\uparrow\rangle)$,
$|D_3\rangle=\frac{1}{\sqrt{3}}(\alpha|d_2\uparrow\rangle+\alpha^2|d_3\uparrow\rangle+|d_4\uparrow\rangle)$ and
$|D_4\rangle=\frac{1}{\sqrt{3}}(\alpha^2|d_2\uparrow\rangle+\alpha|d_3\uparrow\rangle+|d_4\uparrow\rangle)$ with
$j_z=\frac{1}{2},\frac{1}{2},-\frac{1}{2},\frac{3}{2},\frac{1}{2},\frac{1}{2},-\frac{1}{2}$ and $\frac{3}{2}$, respectively.
Another eight molecular orbitals of down spin can be obtained by operating $\mathcal{T}$ on the up spin molecular orbitals:
$\mathcal{T}|P_1\rangle=|p_1\downarrow\rangle$,
$\mathcal{T}|P_2\rangle=\frac{1}{\sqrt{3}}(|p_2\downarrow\rangle+|p_3\downarrow\rangle+|p_4\downarrow\rangle)$,
$\mathcal{T}|P_3\rangle=\frac{1}{\sqrt{3}}(\alpha^2|p_2\downarrow\rangle+\alpha|p_3\downarrow\rangle+|p_4\downarrow\rangle)$,
$\mathcal{T}|P_4\rangle=\frac{1}{\sqrt{3}}(\alpha|p_2\downarrow\rangle+\alpha^2|p_3\downarrow\rangle+|p_4\downarrow\rangle)$,
$\mathcal{T}|D_1\rangle=|d_1\downarrow\rangle$,
$\mathcal{T}|D_2\rangle=\frac{1}{\sqrt{3}}(|d_2\downarrow\rangle+|d_3\downarrow\rangle+|d_4\downarrow\rangle)$,
$\mathcal{T}|D_3\rangle=\frac{1}{\sqrt{3}}(\alpha^2|d_2\downarrow\rangle+\alpha|d_3\downarrow\rangle+|d_4\downarrow\rangle)$ and
$\mathcal{T}|D_4\rangle=\frac{1}{\sqrt{3}}(\alpha|d_2\downarrow\rangle+\alpha^2|d_3\downarrow\rangle+|d_4\downarrow\rangle)$ with
$j_z=-\frac{1}{2},-\frac{1}{2},\frac{1}{2},-\frac{3}{2},-\frac{1}{2},-\frac{1}{2},\frac{1}{2}$ and $-\frac{3}{2}$, respectively.
Each band at the $\Gamma$ point is doubly degenerated (Kramers degeneracy) as shown in the step I of Fig. \ref{evolution}, which
conforms to the band orders in LuSI and YSI exactly. In step II, the exchange coupling of the non-collinear magnetic order is
taken into account, then the time reversal symmetry is broken, and the Kramers degeneracy will split further as shown in the
step II of Fig. \ref{evolution}, which happens to correspond to the case of $\text{GdSI}$. We emphasize that, in both step I
and step II, it is the $|P_2, j_z=\frac{1}{2}\rangle$ ($\mathcal{T}|P_2\rangle, j_z=-\frac{1}{2}$) and
$|D_2, j_z=\frac{1}{2}\rangle$ ($\mathcal{T}|D_2\rangle, j_z=-\frac{1}{2}$) molecular orbitals dominate at the Fermi level
and determine the topological properties, which have the same $j_z$ as the bases studied in our TB model. Therefore, our TB
model discussed in $\mathbf{S1}$ can be used to study the topological properties in LnSI effectively, since they possess the
same symmetry and bases.

 \begin{table*}[tp]
\caption{GGA+SOC calculated total energies of five different magnetic structures for GdSI. The converged magnetic
moments of each Gd atom are given too.}
  \begin{tabular}{l l l l l l l}
\hline
        Config.    &~~~ $\text{Gd}_1$($\mu_B$)          &~~ $\text{Gd}_2$ ($\mu_B$) &~~ $\text{Gd}_3$ ($\mu_B$) &~~$\text{Gd}_4$ ($\mu_B$) &~~Energy (eV) \\
\hline
        FM         &~~~(0, 0, 6.88)&~~	(0, 0, 6.83)	&~~	(0, 0, 6.83)         &~~  (0, 0, 6.83)           &~~-93.826
 \\
        AFM1       &~~~(0, 0, 6.88)&~~	(0, 0, -6.82)	&~~	(0, 0, -6.82)        &~~ (0, 0, 6.82)          &~~-93.818
\\
        AFM2       &~~~(0, 0, 6.88)&~~	(0, 0, 6.82)	&~~	(0, 0, -6.82)        &~~ (0, 0, -6.82)          &~~-93.818
\\
        AFM3       &~~~(0, 0, 6.88)&~~	(0, 0, -6.82)	&~~	(0, 0, 6.82)         &~~  (0, 0, -6.82)         &~~-93.818
\\
        AFM4       &~~~(0, 0, 6.88)&(-$5.91$, -$3.41$, $0.03$) &(0, $6.83$, $0.03$)&($5.91$, -$3.41$, $0.03$)& ~~-93.838
\\
\hline
\end{tabular}
\label{totenergy}
\end{table*}

\section*{S6. MAGNETIC CONFIGURATIONS AND TOTAL ENERGY CALCULATIONS IN GdSI}

\begin{figure}[tp]
\centering
\includegraphics[width=5.5in]{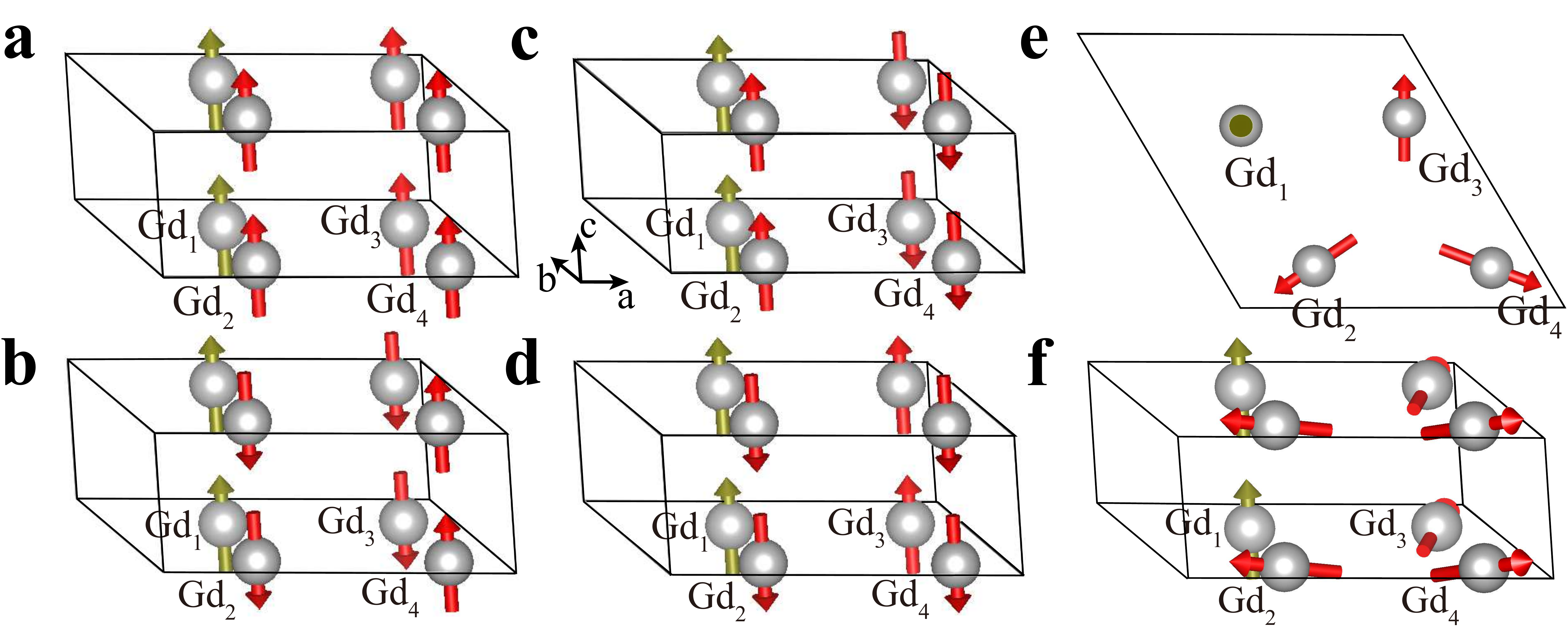}
      \caption{(Color online). Schematic representation of different magnetic configurations for $\text{GdSI}$. (a) FM, (b) AFM1, (c) AFM2 and (d) AFM3. (e) and (f) are top and side views of AFM4, respectively.}
\label{afm}
\end{figure}

\begin{figure}[tp]
\centering
\includegraphics[width=3.in]{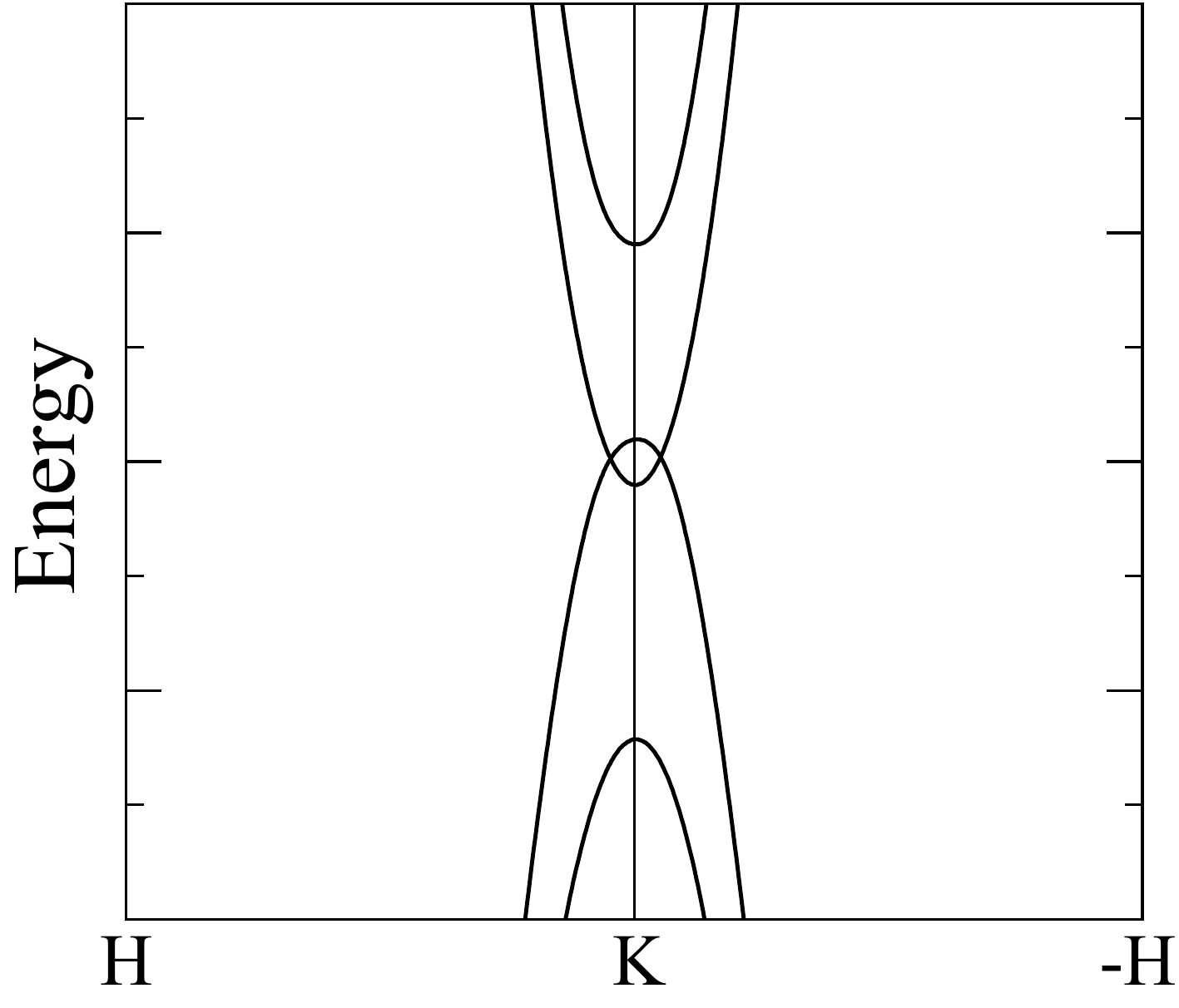}
      \caption{(Color online). GGA+U+SOC calculated band structures along $\text{H}-\text{K}-$-$\text{H}$ of $\text{GdSI}$ for AFM4 magnetic structure.}
\label{mz}
\end{figure}

Five different magnetic structures (FM, AFM1-AFM4) of GdSI are shown in Fig. \ref{afm}, where S and I atoms are omitted for simplicity.
As shown in Fig. \ref{afm}, the magnetic moments of the FM order and three collinear antiferromagnetic (AFM) configurations (AFM1-AFM3)
are aligned along the $c$-direction. Specifically, AFM1 has the AFM exchange coupling along both the $a$- and $b$-directions; AFM2 (AFM3)
only satisfies the AFM exchange coupling along the $a$-direction ($b$-direction), while it has the FM exchange coupling along the
$b$-direction ($a$-direction). For the AFM4 configuration, the magnetic moment of the $\text{Gd}_1$ atom located at the \emph{1c} site
is aligned along the $c$-direction, while the magnetic moments of the \emph{3j} sites $\text{Gd}$ atoms ($\text{Gd}_2$, $\text{Gd}_3$ and
$\text{Gd}_4$) are mainly lying in the $ab$-plane with a $120^\circ$ angle, as shown in Fig \ref{afm}(e), which reduces the magnetic
frustration in the $2\times2$ triangle lattice significantly and leads to the lowest total energy. It is worthy to note that the $C_3$
symmetry is preserved in the AFM4 configuration.

~\\

In order to confirm which is the most favorable magnetic configuration, we have performed the GGA+SOC calculations for all five magnetic
structures, as well as the non-magnetic (NM) state. The calculated total energy of the NM state is about -69.151 $eV$; the total energies
and the converged magnetic moments of the five magnetic structures are summarized in Table \ref{totenergy}. The calculated results show
that the total energy of the NM state is much higher (about 24 eV/u.c.) than that of the magnetic states, indicating that the assumption
of the existence of magnetic order in $\text{GdSI}$ is reasonable. Furthermore, the results listed in Table \ref{totenergy} show that the
three collinear AFM configurations have nearly the same total energy, which are 10 meV higher than that of FM, and 20 meV higher than that
of AFM4. Finally, AFM4 has the lowest total energy among all five magnetic structures, which agrees with our analysis that such magnetic
structure can eliminate the magnetic frustration in the $2\times2$ triangle lattice significantly.

\section*{S7. MIRROR SYMMETRY $M_z$ BREAKING AND FITTED PARAMETERS IN \text{GdSI}}

In addition to the breaking of the time reversal symmetry, the other important effect of the non-collinear AFM structures (AFM4) in GdSI is
the $M_z$ symmetry breaking. Two types of hopping terms can be added to Eq. \ref{H0k} for the breaking of $M_z$ symmetry. (1) We can add a
term that makes the $\pm z$-direction interlayer hopping different, which will mainly result in different band dispersions between
$\pm k_z$-direction, but the energy spectrum in the $k_xk_y$-plane remains unchanged. (2) The existence of the in-plane magnetic moments
means that $S_z$ is not a good quantum number again. Therefore, the nearest intralayer hoppings with the same spin, named as $r_2$, can be
recovered, which will lead to a great change of the band dispersion in the $k_z=0$ plane, but will keep the band energies of the two $k$
points having the same $k_x$, $k_y$ and opposite $k_z$ to be equal. In what follows, we only add such $r_2$ term into the Eq. \ref{H0k}
to study the electronic structures of GdSI, based on the fact that the GGA+U+SOC calculated band dispersions are approximately symmetrical
between $\text{H}-\text{K}$ and $\text{K}-\text{-H}$ directions, as shown in Fig. S6.

~\\

By adding the zeeman splitting and the $r_2$ terms that breaks the $M_z$ symmetry, the total Hamiltonian for \text{GdSI} can be written as following

{\fontsize{10.3}{10}\selectfont
     \begin{align}
       H_{\text{GdSI}}(\boldsymbol{k}) = H(\boldsymbol{k})+
               &\left( \begin{array}{cccc}
           t_A^4                     &                   0         & 0         & 0   \\
                                   0    &-t_A^4                     & 0         & 0 \\
                                   0    &                           0 &  -t_B^4  &       0            \\
                                   0    &                           0 &                0             &   t_B^4
        \end{array}\right)+
        \left( \begin{array}{cccc}
           0                     &                   0         & r_2 G         & 0   \\
                                   0    &0                     & 0         & r_2 G \\
                                   *    &                           * &  0  &       0            \\
                                   *    &                           * &                0             &   0
        \end{array}\right)  \label{Htk} \\
        \nonumber
    \end{align}
}where the definitions of $H(\boldsymbol{k})$ and $G$ are given in the $\textbf{S1}$, and the definitions of $t_A^4$, $t_B^4$ and $r_2$ are given in the main text.
The fitted TB parameters for \text{GdSI} are $t_A^1=2.4116 $ $eV$, $t_A^2=0.022$ $eV$, $t_A^3=0.2619$ $eV$, $t_A^4=0.038$ $eV$,
$t_B^1=3.4477$ $eV$, $t_B^2=-0.0557$ $eV$, $t_B^3=-0.3331$ $eV$, $t_B^4=0.0823$ $eV$, $r_1=0.071$ $eV$, $r_2=0.037$ $eV$ and $\lambda_1=0.012$ $eV$.

\section*{S8. SURFACES STATES OF GdSI}

\begin{figure*}[ht]
\centering
\includegraphics[width=6.in]{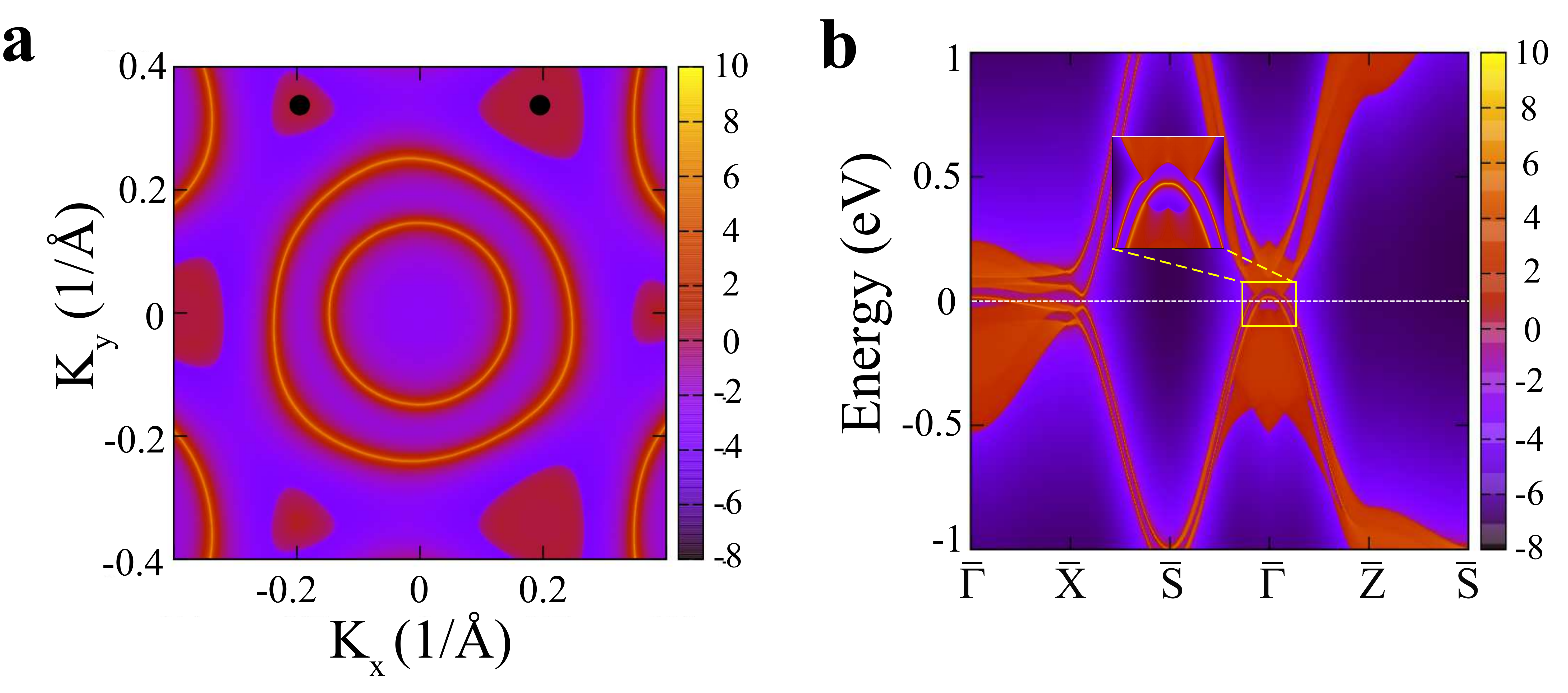}
      \caption{(Color online) (a) Fermi surface for the (001) face of $\text{GdSI}$. The projected positions of the Weyl points on the (001) face are marked by the black dots. (b) (010) surface states of $\text{GdSI}$. The inset shows the surface bands around $\bar{\Gamma}$ point.}
\label{surface}
\end{figure*}

The calculated Fermi surface on the (001) face of $\text{GdSI}$ is shown in Fig. \ref{surface}(a),
which shows that there is no Fermi arc coming out from the projected Weyl points (black dots).
This is because that two bulk WNs carrying opposite chiralities are projected to the same point on
the (001) face. The (010) surface states of $\text{GdSI}$ is plotted in Fig. \ref{surface}(b),
which clearly shows that two non-trivial surface states with opposite fermi velocities connect
the bulk valence and conduction bands along $\bar{S}-\bar{\Gamma}$ and $\bar{\Gamma}-\bar{Z}$
directions (See the inset of Fig. \ref{surface}(b)). It is worth noting that the Fermi circles
around the $\bar{\Gamma}$ point shown in Fig. \ref{surface}(a), as well as the Fermi circle
shown in Fig. 3(d) in the main text, are originated from trivial states. All of them can be
eliminated easily by the surface decoration \cite{wang2016body}.

%

\end{document}